\documentclass[twocolumn]{aa} % para ter duas colunas
\usepackage{graphicx}
\begin{document}

  \title{An estimate of the time variation of the abundance gradient from 
   planetary nebulae\thanks{Based on observations made at the European Southern 
                 Observatory (Chile) and Laborat\'orio Nacional de Astrof\'\i sica 
                 (Brazil)}}

   \subtitle{III. O, S, Ar, and Ne: A comparison of PN samples}

   \author{W. J. Maciel,
          L. G. Lago,
          \and
          R. D. D. Costa
}

   \offprints{W. J. Maciel}

   \institute{Instituto de Astronomia, Geof\'\i sica e Ci\^encias
                 Atmosf\'ericas (IAG), Universidade de S\~ao Paulo - 
                 Rua do Mat\~ao 1226; 05508-900, S\~ao Paulo SP; Brazil\\
                 \email{maciel@astro.iag.usp.br, leonardo@astro.iag.usp.br, 
                 \\roberto@astro.iag.usp.br}
                         }

   \date{Received ; accepted }

   \abstract{The time behaviour of the radial abundance gradients 
   in the galactic disk is investigated on the basis of four different 
   samples of planetary nebulae, comprising both smaller, homogeneous 
   sets of data, and larger, albeit non-homogeneous samples. Four different 
   chemical elements are considered, namely, oxygen, sulphur, argon
   and neon. Our analysis support our earlier conclusions that,
   on the average, the radial abundance gradients have flattened 
   out in the last 6 to 8 Gyr.
   }

   \authorrunning {Maciel et al.}
   \titlerunning {Time variation of radial abundance gradients III.}

   \maketitle
%

%oooooooooooooooooooooooooooooooooooooooooooooooooooooooooooooooooooooooooo
\section{Introduction}
%oooooooooooooooooooooooooooooooooooooooooooooooooooooooooooooooooooooooooo

One of the most interesting observational properties that can be devised 
in order to constrain chemical evolution models is the magnitude as well 
as the space and time variations of the radial abundance gradients in the 
Galaxy (see for example Maciel \& Costa  \cite{mc03}, Maciel 
\cite{maciel00}, and Henry \& Worthey \cite{henry} for recent reviews). 
Regarding the time variation of the gradients, which is possibly the most
important information that can be obtained from the abundance  variations 
in the galactic disk, the galactic subsystem of planetary nebulae (PN) 
plays a particularly important role. These objects have relatively well 
determined chemical compositions, which are essentially based on the 
analysis of bright emission lines, and are originated from stars within 
a reasonably large mass - and therefore age - bracket while on the main 
sequence. As a conclusion, PN can be used to investigate the possibility 
of a time variation of the average gradient in the galactic disk, in the
same way as the open  cluster stars (see for example Friel \cite{friel95}, 
Friel et al. \cite{friel02}).

In the first paper of this series, Maciel et al. (\cite{mcu2003}, 
hereafter referred to as Paper~I) argued in favour of a time flattening of 
the O/H gradient from roughly $-$0.11 dex/kpc to $-$0.06 dex/kpc during the 
last 9~Gyr, or from $-$0.08 dex/kpc to $-$0.06 dex/kpc in the last 5 Gyr, 
on the basis of a large sample of PN in the galactic disk. More recently, 
Maciel et al. (\cite{mlc2005a}, Paper~II) extended the original discussion 
by (i) including S/H data, (ii) adopting  [Fe/H] $\times$  O/H and [Fe/H] 
$\times$ S/H conversions  so that an estimate of the corresponding [Fe/H] 
gradient could be made, (iii) estimating the average gradient from Cepheid 
data and (iv) taking into account some recent determinations of the 
gradients from young objects, such as HII regions and stars in OB 
associations, and presenting a general comparison of the derived gradients 
of all objects considered. It has been shown in Paper~II that the derived 
results lead to a consistent interpretation of a time flattening of the 
radial abundance gradients in the galactic disk, in agreement with the main 
results of Paper~I.

In the present paper, we extend our previous results by taking two 
additional steps: (i) we consider four chemical elements in the discussion, 
namely oxygen, sulphur, argon and neon, so that the ratios O/H, S/H, Ar/H 
and Ne/H can be studied; (ii) we take into account four different PN 
samples, including three additional sets of data apart from the original 
abundances considered in Papers~I and II. These samples are generally 
homogeneous sets, in the sense that all objects have been analyzed by the 
same group, under similar conditions, which includes data reduction and 
abundance determination. This is particularly important in order to make 
the previous results more robust, since all the large samples presently 
available are not homogeneous, being compiled from a few individual 
homogeneous samples.  As we will show in the next few sections, the general 
conclusions of Papers~I and II are maintained.

%oooooooooooooooooooooooooooooooooooooooooooooooooooooooooooooooooooooooooo
\section{The samples}
%oooooooooooooooooooooooooooooooooooooooooooooooooooooooooooooooooooooooooo

\subsection{The basic sample} 

The basic sample is essentially the same used in Papers~I and II (Maciel et 
al. \cite{mcu2003}, \cite{mlc2005a}), and has been described in detail in 
Paper~I, to which the reader is referred. It is based on an earlier sample 
by Maciel \& K\"oppen (\cite{mk94}) and Maciel \& Quireza (\cite{mq99}), 
with some additional objects presented in Costa et al. (\cite{costa04}). 
This sample covers a large range of galactocentric distances,
roughly from $R = 4$ to $R = 14\,$ kpc, with a stronger concentration
of objects located up to $R = 12\,$ kpc. The main additions to our 
earlier sample are the anticentre objects from Costa et al. (\cite{costa04}),
which are typically located at $R > 10\,$ kpc. Such a range is clearly
wide enough to derive accurate gradients, and in fact constitutes
one of the strong points of using PN to study abundance variations
in the galactic disk.

The total number of planetary nebulae for which reliable abundances of O, 
S, Ar and Ne have been obtained is shown in Table~\ref{npn} for the basic 
sample, as well as for the remaining samples considered in this paper.
These additional samples have approximately the same total galactocentric
distance range as the basic sample, but they are smaller and more
dispersed, with comparatively less objects at $R > 10\,$kpc.

It can be seen that the basic sample is the largest sample considered, 
but it should be kept in mind that it is a compilation, albeit  careful, 
of several different determinations in the literature. Therefore, it 
lacks the homogeneity of smaller samples available, so that it is 
interesting to investigate the extent to which the non-homogeneity degree 
affects the results of our analysis. Unfortunately, no sample is presently 
available that includes at the same time a high degree of homogeneity and 
a size large enough to be considered a complete sample -- e.g. by including 
objects in  the whole range of expected abundances.

\begin{table}
\caption[]{Number of PN in each sample.}
\begin{flushleft}
\begin{tabular}{lrrrr}
\hline\noalign{\smallskip}
Sample  & O/H & S/H & Ar/H & Ne/H  \\
\noalign{\smallskip}
\hline\noalign{\smallskip}
Basic Sample  &  234  &  117  &  114  &  118 \\
Henry         &   81  &   77  &   78  &   78 \\
Perinotto     &  115  &  112  &  107  &   92 \\
IAG/USP       &   71  &   58  &   62  &   18 \\
\noalign{\smallskip}
\hline
\end{tabular}
\end{flushleft}
\label{npn}
\end{table}

The distances of the PN in the basic sample come from Maciel \&
Quireza (\cite{mq99}) and Maciel \& K\"oppen (\cite{mk94}), and are
basically from the statistical distance scale by Maciel (\cite{m84}).
For those nebulae not present in the catalogue of Maciel (\cite{m84}),
or for which only distance limits were available, we have adopted
values from more recent scales, especially Cahn et al. (\cite{cks92})
(see the online material associated with Maciel \& Quireza, \cite{mq99}
and Table~1 of Costa et al., \cite{costa04}). The same sources were
used for the remaining samples. The problem of the distances to PN
and their effect on the abundance gradients has been discussed 
elsewhere (see for example Maciel \& K\"oppen, \cite{mk94}). The
analysis of large PN samples is necessarily based on statistical
distances, which include a much larger number of objects than 
individually determined distances. Moreover, the uncertainties of
the individual distances are often larger than generally assumed, 
and in many cases not very different from those associated with 
the statistical distances. We may conclude that the use of different 
statistical distances may have a large influence on the distance 
of a particular object, but the global effect on a large sample is 
negligible, that is,  the distance distribution is essentially 
unaffected. In other words, the abundance gradients are not sensibly 
affected, as well as their derived time variation. A recent study by 
Maciel \& Lago (\cite{ml2005}) led to a similar conclusion regarding 
PN kinematics, which depends on the PN positions -- hence distances -- 
on the galactic disk. It was found that the use of four different 
statistical distance scales (Maciel \cite{m84}, Cahn et al. 
\cite{cks92}, van de Steene \& Zijlstra \cite{vdsz94}, and Zhang 
\cite{zhang95}) for disk PN leads to a similar rotation curve, 
which is not very different from the Population~I curve derived 
from HII regions  (Clemens \cite{clemens}).

\subsection{The sample by Henry et al. (2004)} 

The first homogeneous sample comes from the work of Henry and 
co-workers (Henry et al. \cite{henry04}), who derived abundances 
of a set of 85 planetary nebulae based on observations at KPNO and 
CTIO. The detailed discussion of each step of the derivation is 
given in a series of previous papers (Kwitter \& Henry \cite{kh2001}, 
Milingo et al. \cite{mkhc2002}a,b, and Kwitter et al. \cite{khm2003}). 
The abundances have been derived in a consistent and homogeneous manner 
and, especially, the sulphur abundances are expected to be particularly 
well determined, as they have included  near-IR lines of [SIII] at 
$\lambda$9069, 9532\AA, apart from the optical lines of [SII] at 
$\lambda$6716, 6731\AA \ and [SIII] at $\lambda$6312\AA.

Some of the objects in the Henry et al. analysis are halo nebulae, and 
have been removed from our sample, as they are not adequate to study 
disk properties such as abundance gradients.

Henry et al. (\cite{henry04}) have also analyzed the existence of radial 
abundance gradients from their sample. Although no attempt has been made 
to investigate the time variation of the gradients, they were successful 
in deriving average gradients in the range $-0.04$ to $-0.05$ dex/kpc 
for all studied elementes, which are O/H, Ne/H, S/H, Cl/H and Ar/H. 
Since their abundances -- and to a large extent their distances -- are 
completely independent from the sample considered in our previous 
papers, the fact that measurable and similar gradients have been detected 
is a nice confirmation of the usefulness of disk planetary nebulae as 
a chemical evolution tool.

\subsection{The sample by Perinotto et al. (2004)}

Recently, Perinotto et al. (\cite{perinotto}) have presented a detailed 
analysis of the chemical composition of a large sample of planetary 
nebulae, for which they have used the same extinction corrections, 
electron temperature, densities and ionic abundance determinations, 
as well as ionization correction factors, including atomic constants 
and nebular models. Since the measured fluxes originated from different 
sources, this is not strictly an entirely homogeneous sample, but its 
degree of homogeneity is clearly much higher than in the case of a 
simple compilation, in view of rigorous criteria adopted for the 
inclusion of a given object and the same procedure applied to the 
measured fluxes. In fact, Perinotto et al. (\cite{perinotto}) consider 
their sample as a highly homogeneous data set on galactic PN, determined 
with realistic uncertainties. Comparison of their abundances with some 
well determined data in the literature, such as in the work by 
Kingsburgh \& Barlow (\cite{kb94}) generally produce a very good 
agreement. This sample is considerably larger, amounting to over  
130 nebulae, some of which have been removed from our analysis since 
they may also belong to the halo or have no accurate distances.

\subsection{The IAG/USP sample} 

The last sample considered in this paper comes from our own data, 
which we will call the IAG/USP sample. This is a highly homogeneous
sample, although relatively small, reaching about 70 nebulae, 
after removing bulge objects and PN in the Magellanic Clouds. The 
observations have been made either at the 1.6 m LNA telescope in 
Brazil or with the 1.52 m ESO telescope in Chile. In both cases, 
Boller and Chivens Cassegrain spectrographs have been used. Details 
on the observations and reduction procedures can be found in the 
original papers (see for example Costa et al. \cite{costa04} and 
references therein).

\subsection{Comparison of PN abundances} 

\begin{figure}
   \centering
   \includegraphics[angle=-90.0,width=8.5cm]{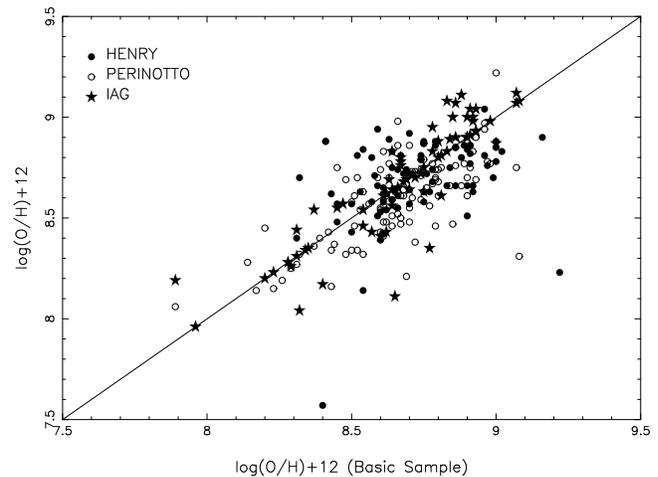}
      \caption{A comparison of the O/H abundances from the 
       samples by Henry et al. (2004, filled circles), Perinotto et al.
       (2004, empty circles) and the IAG/USP sample (stars) with data 
       of the basic sample.}
   \label{pnoh}
   \end{figure}

\begin{figure}
   \centering
   \includegraphics[angle=-90.0,width=8.5cm]{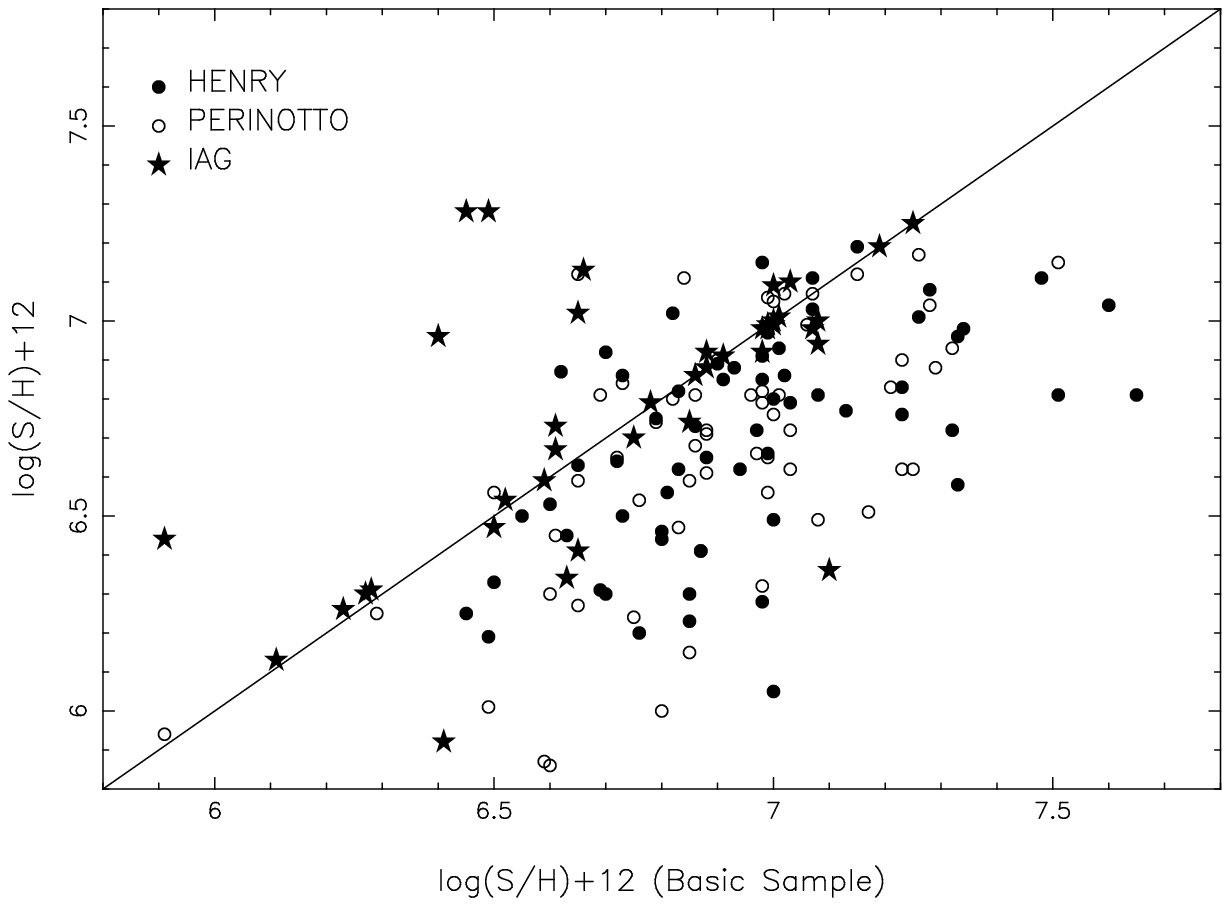}
      \caption{The same as Fig.~\ref{pnoh} for S/H.}
   \label{pnsh}
   \end{figure}

\begin{figure}
   \centering
   \includegraphics[angle=-90.0,width=8.5cm]{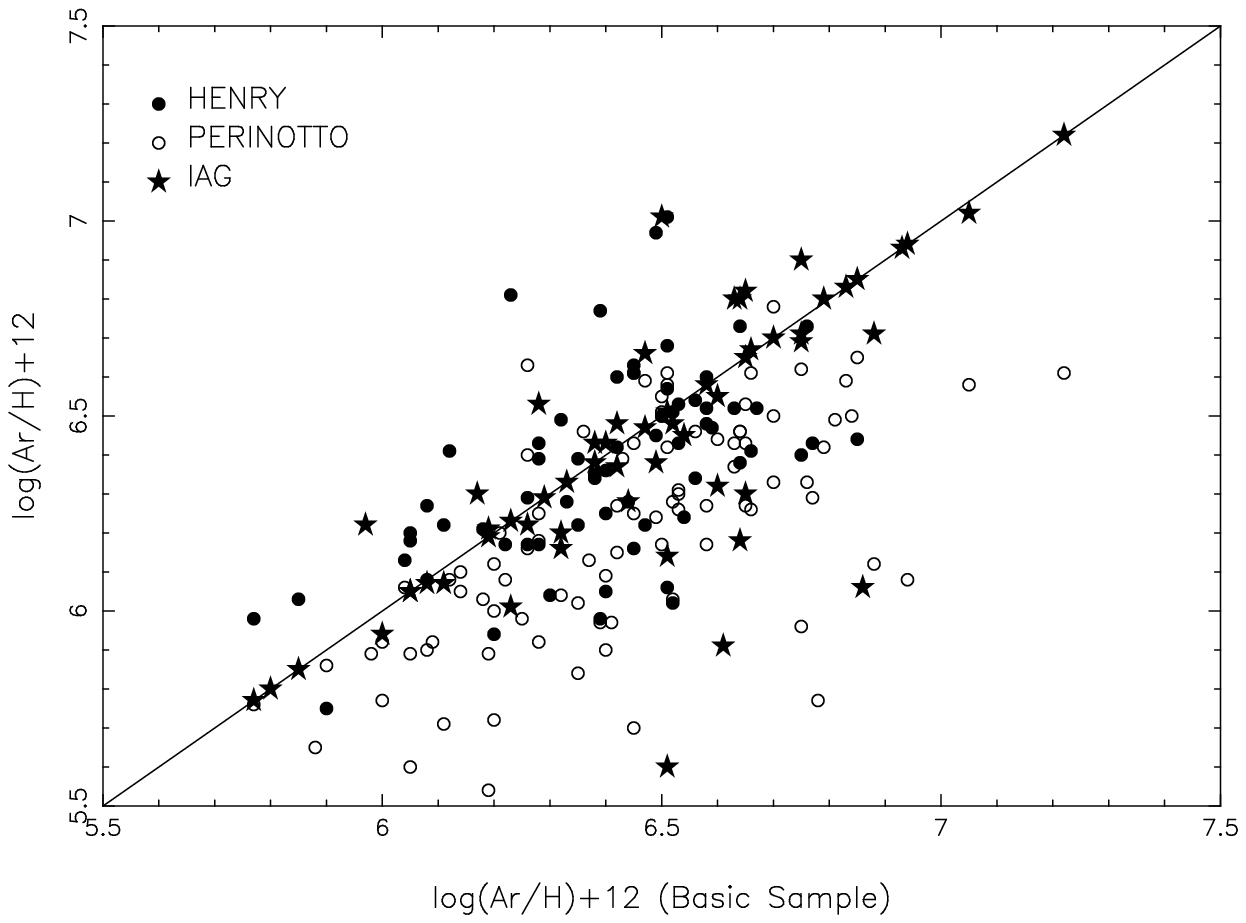}
      \caption{The same as Fig.~\ref{pnoh} for Ar/H.}
   \label{pnarh}
   \end{figure}

\begin{figure}
   \centering
   \includegraphics[angle=-90.0,width=8.5cm]{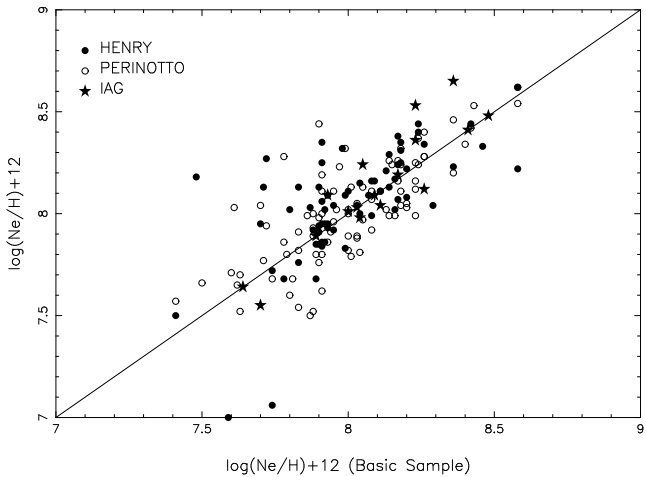}
      \caption{The same as Fig.~\ref{pnoh} for Ne/H.}
   \label{pnneh}
   \end{figure}

A comparison of the abundances of the basic sample and the additional 
samples is shown in Figs.~\ref{pnoh} to \ref{pnneh} for oxygen,
sulphur, argon and neon, respectively. Regarding the sample by Henry 
et al. (\cite{henry04}), (filled circles) it can be seen that the oxygen, 
argon, and neon abundances generally show a very good agreement with the 
basic sample, while part of the sulphur abundances shows some underabundance 
relative to the basic sample. In fact, the lower sulphur abundances have been 
noticed by Henry et al. (\cite{henry04}) when making comparisons with HII 
region data, and was referred to as the \lq\lq sulphur anomaly\rq\rq. 
Several reasons have been suggested to explain such a discrepancy (see 
the discussion by Henry et al.), to which we could add the fact that their 
[SIII] electron temperatures are systematically higher than the 
corresponding [OIII] temperatures, which tends to decrease the derived 
sulphur abundances. A few outliers can be observed in Figs.~\ref{pnoh} to 
\ref{pnneh}, and it should be noticed that some of them, such as Hb 12 
and IC418 have also been identified as  outliers in the plots presented by 
Henry et al. (\cite{henry04}).

The sample by Perinotto et al. (\cite{perinotto}, empty circles) also 
shows a generally good agreement with the data in the basic sample, 
especially for oxygen (Fig.~\ref{pnoh}) and neon (Fig.~\ref{pnneh}). 
Their sulphur abundances, and to a lesser extent, argon, are generally 
lower than in the basic sample (cf. Figs.~\ref{pnsh} and \ref{pnarh}), 
and show the same sulphur anomaly as discussed by Henry et al. 
(\cite{henry04}).

The smaller IAG sample (stars) shows several nebulae directly on the
straight line of Figs.~\ref{pnoh} to \ref{pnneh}, which reflects the fact 
that we have adopted several of the results of this sample in our merged 
basic sample. An exception is neon, for which our own data is very limited, 
as can be seen from Fig.~\ref{pnneh} or Table~\ref{npn}.

%------------------------------------------------------------------------
%------------------------------------------------------------------------

%------------------------------------------------------------------------
%------------------------------------------------------------------------

%------------------------------------------------------------------------
%------------------------------------------------------------------------

%------------------------------------------------------------------------
%------------------------------------------------------------------------

%oooooooooooooooooooooooooooooooooooooooooooooooooooooooooooooooooooooooooo
\section {The method}
%oooooooooooooooooooooooooooooooooooooooooooooooooooooooooooooooooooooooooo

A detailed discussion of the method used in order to estimate the 
abundance gradients at different epochs is given in Papers~I and II 
(Maciel et al. \cite{mcu2003}, \cite{mlc2005a}) to which the reader is 
referred for details. Basically, the abundance gradients in the form 
$d\log({\rm X/H})/dR$ (dex/kpc) were determined for X = O, S, Ar and Ne 
assuming a linear variation of the abundances with the galactocentric 
distance $R$, namely, neglecting possible space variations within the 
galactic disk of the gradients themselves. A value of $R_0 = 8.0\,$kpc 
was adopted for the position of the LSR, as in the previous papers. 

\begin{figure}
   \centering
   \includegraphics[angle=-90.0,width=8.5cm]{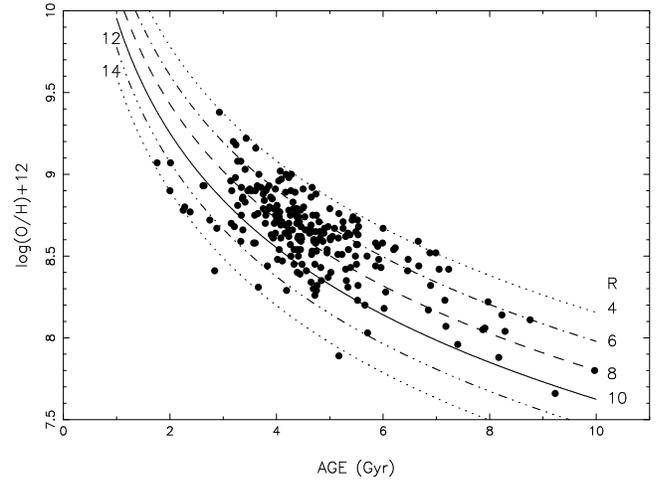}
      \caption{The adopted age-metallicity relation from 
      Edvardsson et al. (\cite{edv93}) which depends on
      the galactocentric distance $R$. The AMR are plotted
      for O/H abundances and galactocentric distances in 
      the range $4 < R ({\rm kpc}) < 14$. The dots show the 
      PN of the basic sample.}
   \label{pnamr}
   \end{figure}

The samples were divided into age groups, and the ages have been estimated 
in the following way. First, the heavy element abundances, that is, O/H, 
S/H, Ar/H and Ne/H, have been converted into [Fe/H] metallicities using
a well determined [O/H] $\times$ [Fe/H] relationship for the galactic
disc, as well as correlations of S/H, Ar/H and Ne/H with the oxygen 
abundance O/H. These correlations have been studied by several people 
(see for example Henry et al. \cite{henry04}), and the slopes are very 
close to unity, reflecting the fact that, as a first approximation, all 
four element ratios are good tracers of the interstellar abundances at 
the time of formation of the PN progenitors. From the [Fe/H] metallicities, 
the ages have been determined using an age-metallicity relationship 
(AMR) by Edvardsson et al. (\cite{edv93}), which also depends on the 
galactocentric distance. Such a relationship constitutes an improvement 
relative to the solar neighbourhood AMR generally adopted, although it 
should be considered as an approximation as applied to the galactic disk. 
As an example, Fig.~\ref{pnamr} shows the adopted age-metallicity relation 
as applied to the O/H abundances of the PN in the basic sample. It can be 
seen that the whole age bracket extends from 1 Gyr to 10 Gyr, approximately, 
with  a larger concentration of objects with ages around 4 to 5 Gyr.

%------------------------------------------------------------------------
%------------------------------------------------------------------------

The adoption of an age-metallicity relation is probably
our main uncertainty source, even allowing for some galactocentric
dependence. Although most average AMR in the literature are
similar to each other (see for instance Rocha-Pinto et al.,
\cite{rp2000}), there is still an ongoing discussion on
their intrinsic dispersion at any given age. An alternative
approach involves the direct determination of the central
star masses, and hence ages, from the nebular chemical
composition, as we have done in Paper~I, with similar results
for the time variation of the abundance gradients. Such a
procedure can also be applied to the predictions of theoretical
models of intermediate mass stars. As an example, we have 
recently considered models by Marigo (\cite{marigo}) and
our preliminary results (Maciel, Costa \& Lago, in preparation)
largely support the results based on the age-metallicity relation.

Once the individual ages have been determined, the nebulae in each 
sample have been divided into two age groups, one called the 
{\it younger} Group~I, and the other the {\it older} Group~II. 
As in Paper~II, if $t_I$ is the upper age limit of Group~I, all PN 
progenitors having ages $t \leq t_I$ will belong to Group~I, while 
those with ages $t > t_I$ belong to Group~II. Several values of 
$t_I$ have been considered in the range 
$3.0  < t_I {\rm (Gyr)} < 6.0$, and for each of these values
we have calculated  the gradients of Groups I and II and the 
corresponding correlation coefficients. Therefore, our conclusions 
do not depend on the adopted age limit $t_I$, that is, they are 
independent of the detailed manner according to which the groups are
defined. Of course, for values of the age limit very close to the 
limiting values of the adopted range, namely $t_I \simeq 3$ or 6 Gyr, 
either Group~I or Group~II becomes underpopulated, so that the 
corresponding results are less reliable (see Paper~II for a 
detailed discussion).

It should be stressed that the main goal of this project is to establish
the {\it time variation} of the abundance gradients, that is, to
determine whether the gradients steepen or flatten out with time.
Therefore, we are not particularly concerned with an accurate
determination of the {\it absolute magnitude} of the gradients
at any given epoch.

%oooooooooooooooooooooooooooooooooooooooooooooooooooooooooooooooooooooooooo
\section{Results and discussion}
%oooooooooooooooooooooooooooooooooooooooooooooooooooooooooooooooooooooooooo

\subsection{O/H} 

\begin{figure*}
   \centering
   \includegraphics[angle=-90.0,width=14.5cm]{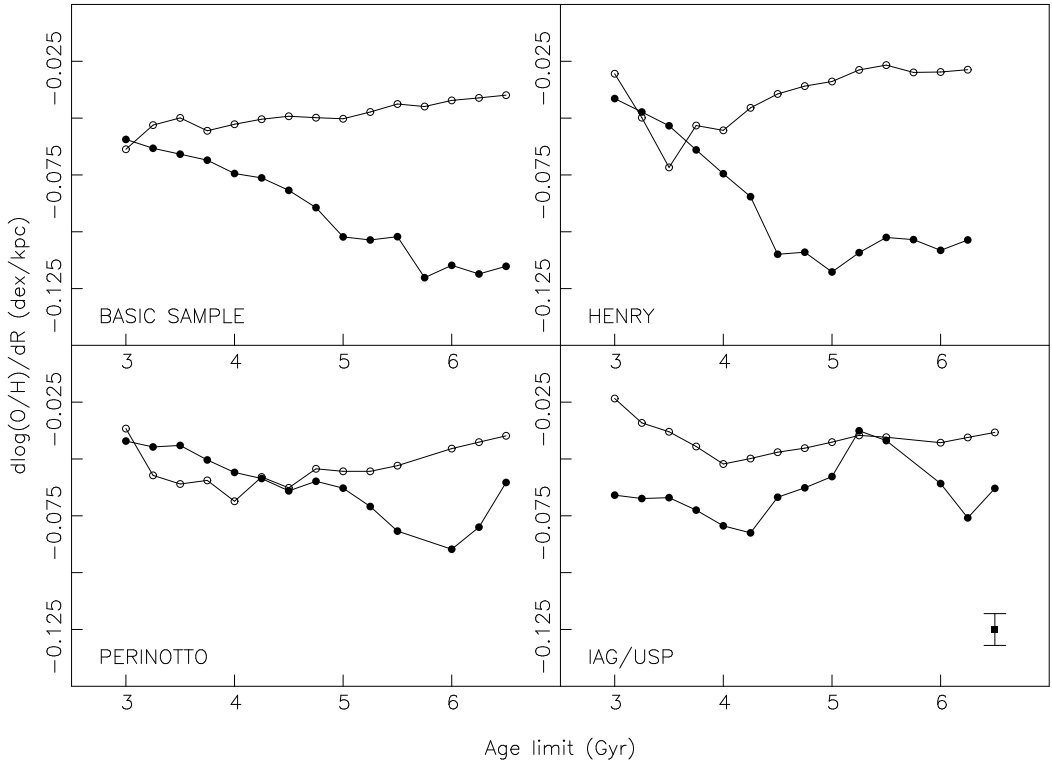}
      \caption{Time variation of the O/H gradient from planetary
      nebulae. The four PN samples were divided into two
      age groups, Group I (\lq\lq younger\rq\rq), with ages lower
      than the age limit $t_I$, and Group II (\lq\lq older\rq\rq), 
      with ages higher than $t_I$. The plot shows the O/H gradient 
      (dex/kpc) of each group as a function of the upper age 
      limit of Group~I, $t_I$, for the four samples considered. 
      The gradients of the younger Group~I (empty circles connected by 
      lines) are flatter than those of the older Group~II 
      (filled circles connected by lines). An average error bar
      is given at the lower right corner of the last panel.}
   \label{oh4}
   \end{figure*}

The main results on the time variation of the abundance gradients of the 
O/H ratio are shown in Fig.~\ref{oh4}, where we plot the derived gradients
as a function of the adopted age limit $t_I$. These are the best results of all 
four elements, as the average O/H abundances are higher, better determined 
and the samples are larger. From Fig.~\ref{oh4}, it can be concluded that 
the gradients of the younger Group~I (empty circles connected by lines) 
are systematically flatter than the corresponding gradients of the older 
Group~II (filled circles connected by lines), supporting our earlier 
conclusions of Papers~I and II. These results apply to all four samples 
considered, especially the larger basic sample, the sample by Henry et 
al. (\cite{henry04}) and the IAG/USP sample. For the basic sample (top 
left panel) the results are essentially the same as presented in Fig.~1 
of Paper~II.

%------------------------------------------------------------------------
%------------------------------------------------------------------------

In the case of the sample by Perinotto et al. ({\cite{perinotto}, bottom 
left panel), the gradients of both groups are similar for age limits in 
the range $3 < t_I ({\rm Gyr}) < 4.5$, but for larger age limits the 
separation of the two groups is clear. 

In view of the comment at the end of Sec.~3, for $t_I \simeq 3\,$Gyr the 
younger Group~I is underpopulated, and the same occurs at 
$t_I \simeq 6\,$Gyr for Group~II. Therefore, near these limiting values 
the statistical results are less reliable, as the space distribution
of the nebulae may not be homogeneous. We would then expect the 
best results to be associated with groups containing a reasonably large 
fraction of the whole sample. As an illustration, the age limit at which 
both groups have the same number of elements (that is, half of the sample) 
occurs typically for $3.9 < t_I ({\rm Gyr}) < 4.6$, showing no systematic
differences between the space distributions of the PN groups.

A detailed discussion of the errors involved in the determination of
the gradients, as well as a statistical analysis of the different
object samples considered have been recently given by Maciel et al.
(\cite{mlc2005b}). For the sake of completeness, we have included at 
the lower right corner of the last panel of Fig.~\ref{oh4} an average
error bar concerning the O/H gradient from PN. The same procedure
has been adopted in the corresponding figures of the other element
ratios, namely S/H, Ar/H and Ne/H.

Another important quantity that can be used in the interpretation of 
Fig.~\ref{oh4} is the correlation coefficient of the linear fits. For O/H, 
the coefficients of all fits shown in Fig.~\ref{oh4} are large, typically 
in the range $0.50 < \vert r\vert  < 0.90$, so that the flattening of the
gradients of the younger Group~I relative to the older Group~II is real.

\subsection{S/H} 

For the S/H ratio, the same behaviour of the O/H gradients is observed, 
as can be seen from Fig.~\ref{sh4}. This is particularly true for the 
basic sample, the sample by Henry et al. (\cite{henry04}) and the 
IAG/USP sample. Again the gradients of the younger Group~I are 
systematically flatter than for the older Group~II, and the correlation 
coefficients are also generally large, in the range 
$0.50 < \vert r\vert  < 0.80$. For the basic sample, in particular, the 
correlation coefficients of both groups are large, 
$0.70 < \vert r\vert  < 0.80$. The results of the top left panel are 
also essentially the same as in Fig.~2 of Paper~II. For these three 
samples, the average gradients of Group~I are essentially independent 
of the adopted age limit, while the corresponding gradients of the older 
groups are increasingly steep, although the derived values for age limits 
close to the upper limit of 6 Gyr are less reliable, as mentioned.

%------------------------------------------------------------------------
   \begin{figure*}
   \centering
   \includegraphics[angle=-90.0,width=14.5cm]{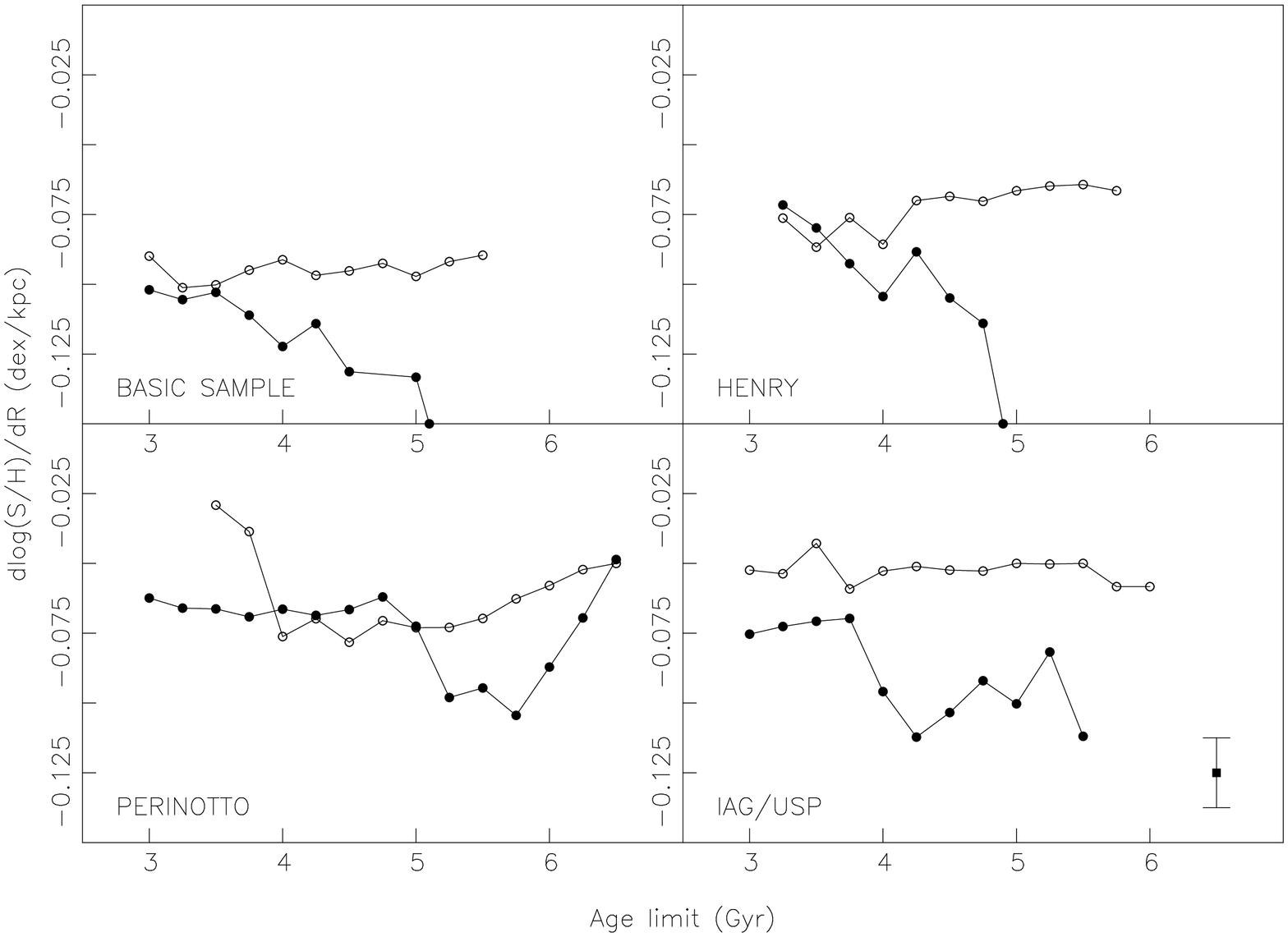}
      \caption{The same as Fig.~\ref{oh4} for S/H.}
   \label{sh4}
   \end{figure*}
%------------------------------------------------------------------------

It is interesting to notice that the behaviour of S/H gradients of the
basic sample and the Henry et al. sample are very similar, as shown by
the top panels of Fig.~\ref{sh4}. As mentioned in Sec.~2, the sulphur
abundances of the sample by Henry et al. are expected to be very
accurate, providing a further confirmation of the flattening of the
gradients for the younger Group~I. It can also be concluded that
the \lq\lq sulphur anomaly\rq\rq\ present in the Henry et al. sample
does not affect the picture of the time evolution of the gradients
as derived from the basic sample. 

In the sample by Perinotto et al. (\cite{perinotto}), S/H shows a similar 
behaviour as O/H, in the sense that the gradients of both groups are 
of the same order of magnitude for smaller values of the age limit, 
$t_I < 4.5\,$Gyr, presenting some flattening of the younger Group~I at 
higher values of $t_I$. For this sample, the correlation coefficients 
reach the lowest values, typically $\vert r\vert \simeq\,$ 0.4 to 0.5. 
Again for illustration purposes, in all cases of Fig.~\ref{sh4}, both 
Groups~I and II are approximately of the same size for 
$4.1 < t_I ({\rm Gyr}) < 4.6$.

\subsection{Ar/H} 

The Ar/H gradients shown in Fig.~\ref{arh4} also indicate some flattening
of the younger Group~I in the basic sample, especially for age limits 
$t_I > 4.5$ Gyr. The corresponding correlation coefficients are moderate, 
typically $\vert r\vert \simeq 0.50$.

%------------------------------------------------------------------------
   \begin{figure*}
   \centering
   \includegraphics[angle=-90.0,width=14.5cm]{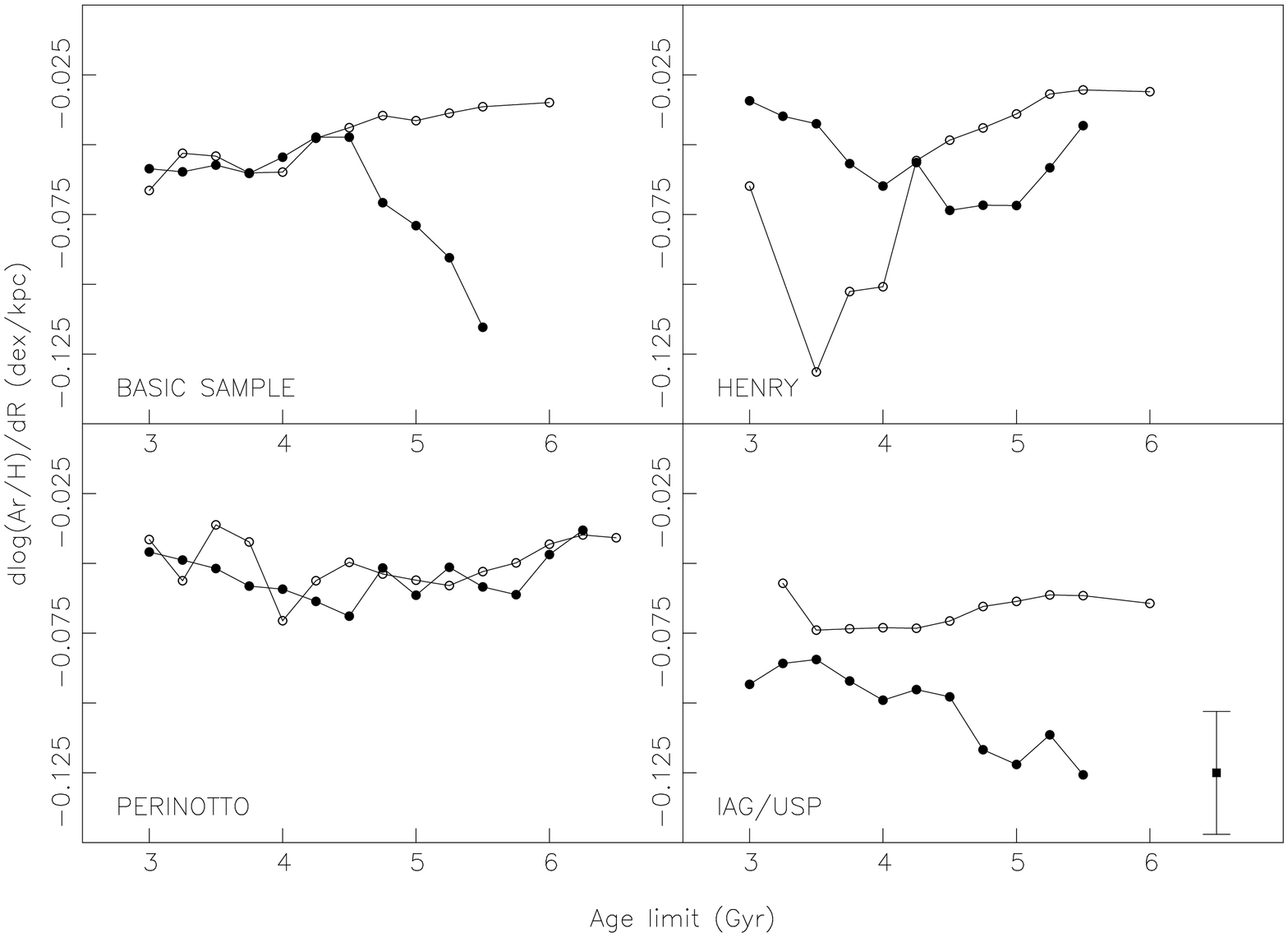}
      \caption{The same as Fig.~\ref{oh4} for Ar/H.}
   \label{arh4}
   \end{figure*}
%------------------------------------------------------------------------

For the IAG/USP sample, the flattening is more evident, as can be seen from 
the bottom right panel of Fig.~\ref{arh4}, where Group~I (empty circles) 
indicate less pronounced gradientes than Group~II (filled circles). Here 
the correlation coefficients are higher, in the range 
$0.50 < \vert r\vert  < 0.80$. 

In the sample by Henry et al., some flattening of Group~I can also be 
noticed for $t_I > 4\,$Gyr. In all panels of Fig.~\ref{arh4} both groups 
have similar sizes for $3.8 < t_I ({\rm Gyr}) < 4.6$, so that the 
observed flattening is probably real. However, the correlation 
coefficients in this range are modest, typically 
$\vert r \vert \simeq 0.40$ for Group~I and somewhat higher 
($\vert r \vert \simeq 0.60$) for Group~II, so that this conclusion 
cannot be stressed too much. 

For this element, the sample by Perinotto et al. does not lead to any 
definite conclusions. In this case, the gradients of both groups are 
very similar, within the expected uncertainties of the gradients 
themselves, which are typically 0.02 dex/kpc, as can be seen from 
Papers~I and II. Since the correlation coefficients are modest
($\vert r \vert \simeq 0.4$ to 0.6), we should probably conclude that 
no differences can be observed between the gradients of the different 
groups. 

\subsection{Ne/H} 

Finally, Fig.~\ref{neh4} shows the results of the Ne/H ratio for our 
adopted samples. We can see that a well defined flattening of the younger 
Group~I relative to the older Group~II is apparent for the Henry et 
al. (\cite{henry04}) sample, for which the correlation coefficients 
are also moderate to large, being in the range
$0.40 < \vert r\vert  < 0.60$ for Group~I and
$0.50 < \vert r\vert  < 0.80$ for Group~II. Such a result is particularly 
important, as the remaining two samples do not allow clear conclusions 
to be drawn. For this sample, both groups have approximately the same
size for $t_I \simeq 4.3\,$Gyr.

%------------------------------------------------------------------------
   \begin{figure*}
   \centering
   \includegraphics[angle=-90.0,width=14.5cm]{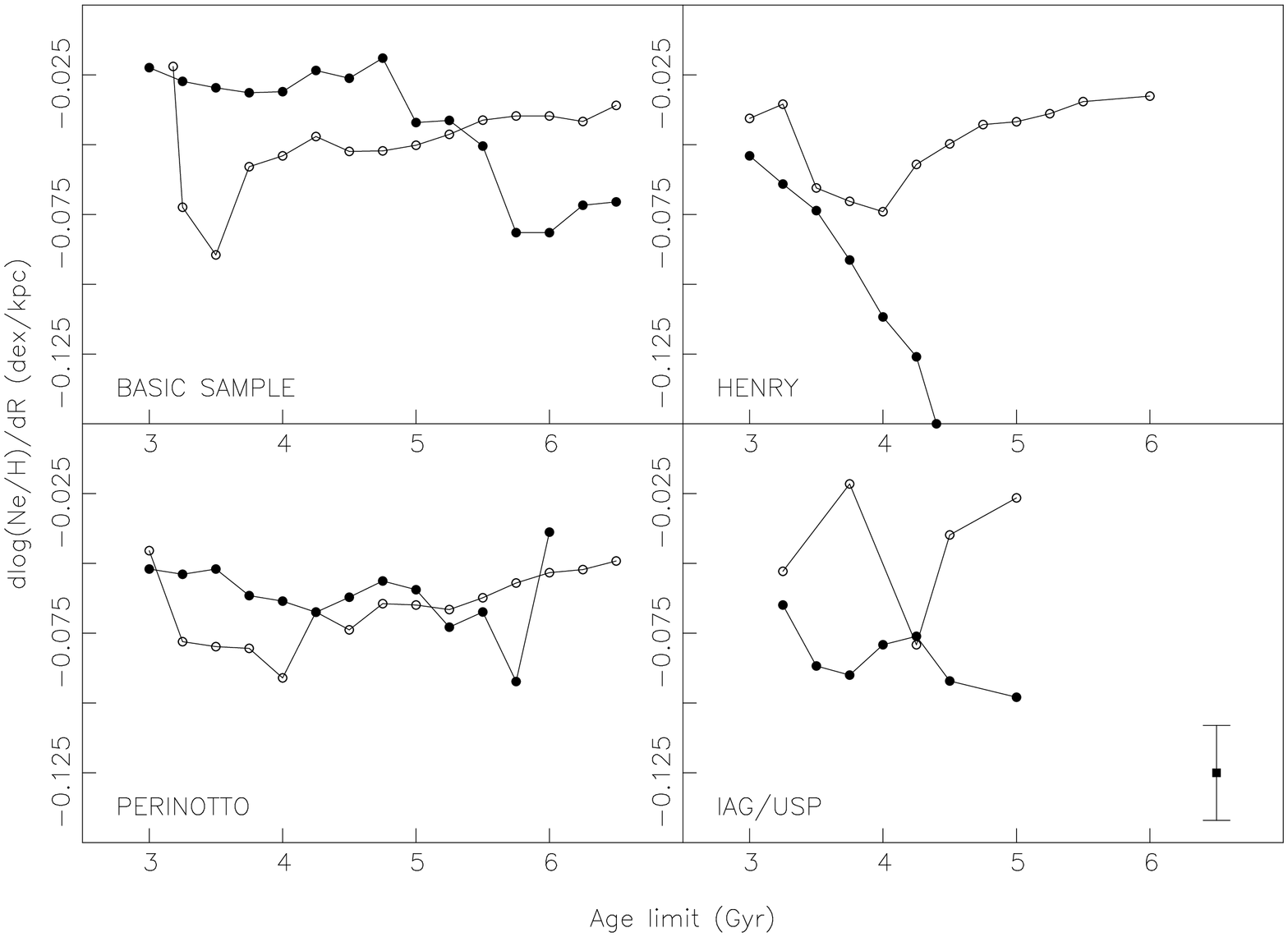}
      \caption{The same as Fig.~\ref{oh4} for Ne/H.}
   \label{neh4}
   \end{figure*}
%------------------------------------------------------------------------

The Perinotto et al. sample displays essentially similar gradients for 
both age groups. The correlation coefficients are large in this case, 
typically in the range $0.50 < \vert r\vert  < 0.80$, so that we conclude 
from this sample that the gradients of both groups are similar.

In the case of the basic sample, acceptable gradients can be obtained 
only for the younger Group~I, for which the correlation coefficients 
are well defined, and typically in the range $0.40 < \vert r\vert  < 0.80$,
so that the derived  gradients are probably correct. However, for 
the older groups the gradient is very flat for most of the range in 
age limits, and the correlation coefficients are very low, typically 
$\vert r \vert < 0.20$, which is consistent with the very low gradients 
displayed by this group. This means that the gradients are not well 
defined in this group, so that no comparison of the time evolution of 
the gradients can be made.

As mentioned before, the IAG/USP sample includes only a very limited 
number of PN with well determined Ne/H abundances, so that no reliable 
gradients can be obtained from this sample. The corresponding panel in 
Fig.~\ref{neh4} shows some indication of a flattening for the younger 
Group~I, but the correlations are rather poor and the sample is very 
small. The gradients of the older Group~II are better, so that we can 
conclude the results of this sample are consistent with the results
of the sample by Henry et al. It seems that for this element the 
homogeneity of the sample is especially important, as the the less 
homogeneous samples (the basic and the Perinotto samples) present
uncertain and even conflicting results, while the the homogeneous 
sample by Henry et al. (\cite{henry}) shows very clear results.

%oooooooooooooooooooooooooooooooooooooooooooooooooooooooooooooooooooooooooo
\section{Time evolution of the [Fe/H] gradient} 
%oooooooooooooooooooooooooooooooooooooooooooooooooooooooooooooooooooooooooo

Taken together, the results shown in Figs.~\ref{oh4} to \ref{neh4}
generally confirm the main conclusions of Papers~I and II, in the
sense that there is a clear tendency for the gradients to flatten out, 
at least during the lifetimes of the older objects considered here. 
Adopting 13.6 Gyr for the age of the Galaxy, as in the previous papers, 
the gradients have been flattening out in the last 6 to 8 Gyr, 
approximately, or since the Galaxy was approximately 6 Gyr old.
From the data considered here, as in the previous papers, nothing
can be said about the time evolution of the abundance gradients
during the early epochs of galactic evolution, that is, for
$t \leq 6\,$Gyr. These results cannot be taken as definitive,
in view of the uncertainties and assumptions considered, especially
regarding the abundances and age determinations. However, as we
will see below, we have considerably expanded our original analysis
to include new elements, samples and galactic objects, so that
our conclusions are now based on a more solid foundation.

The inclusion of Ar and Ne generally supports the earlier conclusions
drawn from O/H (Paper~I) and S/H (Paper~II), although these
elements are clearly more limited in view of the lower, more 
uncertain abundances, and the smaller samples. In this respect,
The homogeneous sample by Henry et al. (\cite{henry04}) is 
particularly useful, especially for neon. Adopting this
sample as representative of the the Ne/H gradient, we would
estimate for Group~I (ages between 0 and 4 Gyr) an average
gradient of $d\log ({\rm Ne/H})/dR \simeq -0.06\,$dex/kpc, and
for Group~II (ages between 4 and 8 Gyr) an average
gradient of $d\log ({\rm Ne/H})/dR \simeq -0.11\,$dex/kpc, which
would correspond approximately to an [Fe/H] gradient 
$d[{\rm Fe/H}]/dR \simeq -0.07\,$dex/kpc and
$d[{\rm Fe/H}]/dR \simeq -0.13\,$dex/kpc for Groups~I and II,
respectively (see Paper~II for details in the conversion of
observed abundance gradients to [Fe/H] gradients).

For Ar/H, using the same procedure, and taking into account
the basic sample as well as the sample by Henry et al. (\cite{henry04}),
we would obtain $d\log ({\rm Ar/H})/dR \simeq -0.04\,$dex/kpc 
or $d[{\rm Fe/H}]/dR \simeq -0.05\,$dex/kpc for Group~I and
$d\log ({\rm Ar/H})/dR \simeq -0.075\,$dex/kpc 
or $d[{\rm Fe/H}]/dR \simeq -0.09\,$dex/kpc for Group~II.

%------------------------------------------------------------------------
   \begin{figure*}
   \centering
   \includegraphics[angle=-90.0,width=13.0cm]{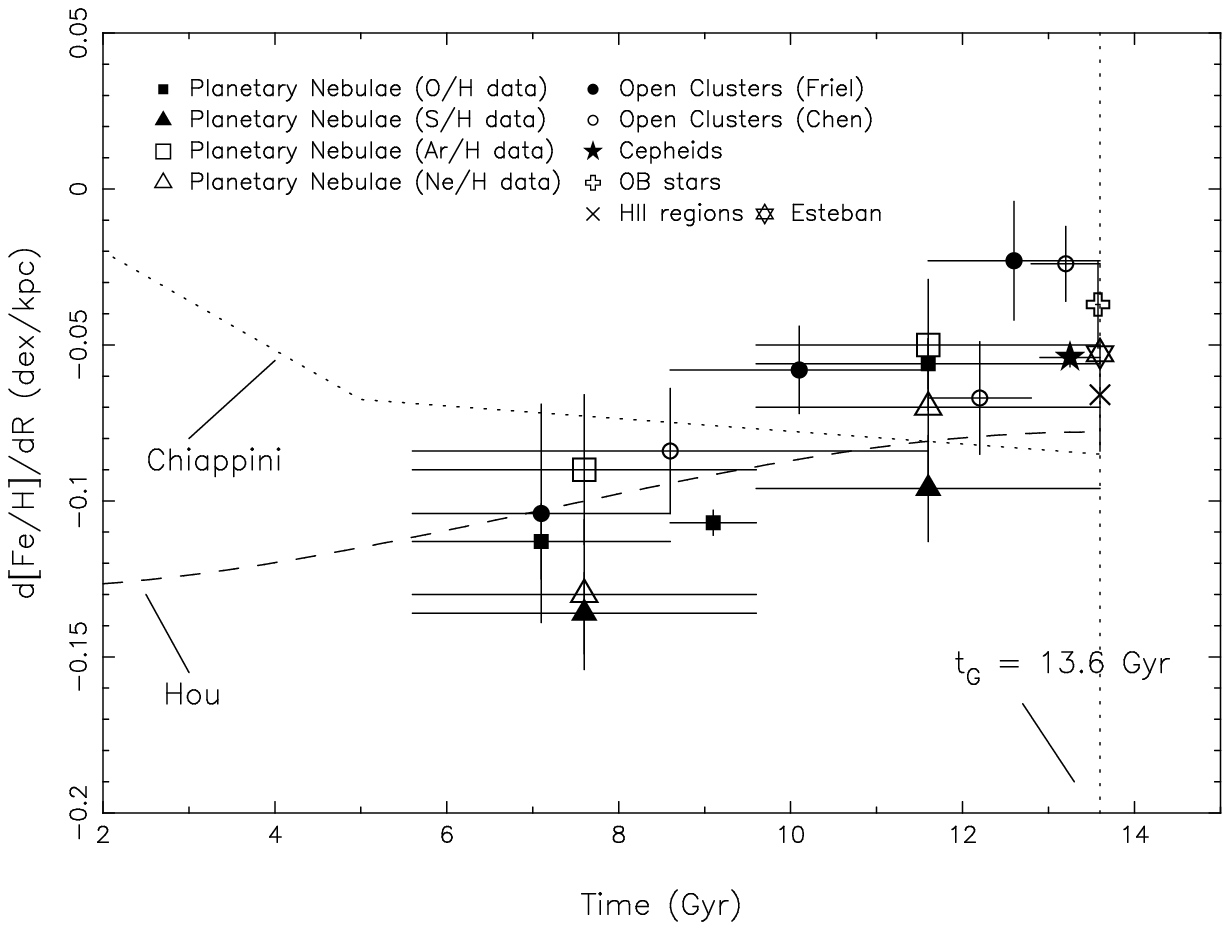}
      \caption{Time variation of the [Fe/H] abundance gradient (dex/kpc). 
      The sme as Fig.~8 from Paper~II, with the following additions: 
      (i) Converted [Fe/H] gradients from Ar/H abundances in PN analyzed 
      in the present paper (empty squares); (ii) the same for Ne/H
      gradients (open triangles); (iii) recent results by Esteban et al. 
      (\cite{esteban}) for HII regions (star of David), and 
      (iv) theoretical tracks from Chiappini et al. (\cite{cmr2001}, 
      dotted line). The remaining data are: (i) [Fe/H] gradients from PN 
      calculated from O/H data (filled squares), (ii) the same for S/H 
      (filled triangles); (iii) Open cluster data from Friel et al. 
      (\cite{friel02}, filled circles); (iv) the same from Chen et al. 
      (\cite{chen}, empty circles); (v) Cepheid data (filled star); 
      (vi) OB stars in associations (thick cross), (vii) HII regions 
      from the literature (x sign), and (viii) theoretical models by Hou 
      et al. (\cite{hou}, dashed line). The dotted vertical line shows 
      the adopted age of the galactic disk, $t_G = 13.6$ Gyr.}
   \label{vargradx}
   \end{figure*}
 %------------------------------------------------------------------------

These estimates are plotted in Fig.\ref{vargradx}, which reproduces
Fig.~8 of Paper~II with the following additions: (i) the Ar/H gradients 
from PN converted into [Fe/H] gradients are included, as described above 
(empty squares); (ii) the same for Ne/H gradients (open triangles);  
(iii) recent results by Esteban et al. (\cite{esteban}) for 
O/H gradients in HII regions based on abundances derived from recombination 
lines, converted into [Fe/H] gradients as in Paper~II (star of David), 
and (iv) predictions of theoretical models (model A) by Chiappini et al. 
(\cite{cmr2001}) are shown as a dotted line. The remaining results plotted 
in Fig.~\ref{vargradx}  are the same as in Fig.~8 of Paper~II, namely: 
(i) O/H data from PN  converted into [Fe/H] gradients, considering
three age groups (young, intermediate, old) (filled squares); 
(ii) the same for S/H (filled triangles); (iii) [Fe/H] data from open 
clusters from the sample by Friel et al. (\cite{friel02}, filled circles); 
(iv) the same for the sample by Chen et al. (\cite{chen}, open circles); 
(v) [Fe/H] data from cepheids (Andrievsky et al. \cite{serguei1}abc,
\cite{serguei5}, Luck et al. \cite{serguei4}, filled star); (vi) OB stars and
associations from Daflon \& Cunha (\cite{daflon}, thick cross), and 
(vii) HII region data from the recent literature, as discussed in 
Paper~II (x sign). Finally, the dashed line shows predictions by 
theoretical models by Hou et al. (\cite{hou}), which can be compared
with the previously mentioned results by Chiappini et al.
(\cite{cmr2001}) as illustrations of theoretical models of chemical
evolution. The dotted vertical line indicates the adopted age of the 
galactic disk, $t_G = 13.6$ Gyr. 

As mentioned in Paper~II, our purpose here is to provide observational
constraints to chemical evolution models, rather than to discuss in
detail the different models available in the literature. Nevertheless,
it is interesting to comment on the large discrepancy of the models
by Hou et al. (\cite{hou}) and Chiappini et al. (\cite{cmr2001}) as 
observed in Fig.~\ref{vargradx}, especially at earlier epochs of the
galactic lifetime. Hou et al. (\cite{hou}) adopt an exponentially 
decreasing infall rate and an \lq\lq inside-out\rq\rq\ scheme for the
formation of the galactic disk, in which a rapid increase of the metal
abundance at early times in the inner disk leads to a steep gradient.
As the star formation migrates to the outer disk, metal abundances
are enhanced in that region, so that the gradients flatten out. Similar
results have also been obtained by Alib\'es et al. (\cite{alibes}),
among others. In the models by Chiappini et al. (\cite{cmr2001}), two
infall episodes are assumed to form the halo and the disk. The latter
also has an \lq\lq inside-out\rq\rq\ formation scenario, in which the
timescale is a linear function of the galactocentric distance. The
detailed time variation of the gradients depends on the particular
model, but in general some steepening of the gradients is predicted,
which does not seem to be supported by the data presented in 
Fig.~\ref{vargradx}. Possibly, the different timescales for star formation
and infall in the models by Hou et al. (\cite{hou}) and Chiappini et
al. (\cite{cmr2001}) are responsible for their different predictions
on the time variation of the abundance gradients.

It can also be seen that the new results fit nicely the picture previously 
sketched in Paper~II, so that the same average flattening rate can be advocated 
here, namely,  $d[{\rm Fe/H}]/dR \sim 0.005$ to $-0.010\,$dex kpc$^{-1}$ Gyr$^{-1}$ 
for the last 6 to 8 Gyr. This value refers to a spatially-averaged 
gradient, corresponding to a linear variation of the abundances with
the galactocentric distance. In practice, the situation is more 
complicated, and all quoted theoretical models predict some space
variations of the gradients along with their temporal variations. Since
the observational results on the space variations are still very limited
and often contradictory, further work is needed in order to definitely
establish the characteristics of this important observational constraint.

\begin{acknowledgements}
      This work was partly supported by  FAPESP (grant 02/08816-5), 
      CNPq and CAPES.  Observations at ESO/Chile were possible through the 
      FAPESP grant 98/10138-8.
      
\end{acknowledgements}

%oooooooooooooooooooooooooooooooooooooooooooooooooooooooooooooooooooooooooo

\end{document}